\documentstyle[12pt]{article}

  \def\ba{{\mbox {\boldmath $a$}}}
  \def\bb{{\mbox {\boldmath $b$}}}
  \def\be{{\mbox {\boldmath $e$}}}
  \def\bn{{\mbox {\boldmath $n$}}}
  \def\bJ{{\mbox {\boldmath $J$}}}

  \def\fbea{\mbox {\footnotesize{$\be^{(\alpha)}$}}}
  \def\fbJ{\mbox {\footnotesize{$\bJ$}}}
  \def\fbn{\mbox {\footnotesize{$\bn$}}}

\begin{document}

\noindent
\bigskip
\bigskip

\begin{center}

{\large\bf A Phenomenological Formula for KM Matrix}\\[1in]

{\bf Kouzou Nishida and Ikuo S. Sogami }\\[.2in]

{\it Department of Physics, Kyoto Sangyo University, \\
Kita-Ku, Kyoto 603, Japan}\\[.2in]

August 17, 1996  \\[.8in]

{\large\bf Abstract}\\[.1in]
\end{center}
\begin{quotation}
   We propose a phenomenological formula relating the Kobayashi-Maskawa
  matrix $V_{KM}$ and quark masses in a form $(m_d,\ m_s,\ m_b)\propto
  (m_u,\ m_c,\ m_t)V_{KM}$.  The formula agrees with experimental data well
  and has an interesting geometric picture.  The origin of such a formula
  is discussed in the standard model.
\end{quotation}

\newpage

  Understanding the origin of fermion masses and the Kobayashi-Maskawa (KM)
matrix [1] is one of the major problems awaiting solution in particle
physics. In the standard model, all components of those matrices are free
parameters which must be adjusted by experiment. Until now many
attempts [2--9] have been devoted to find formulas relating the fermion
masses and the KM matrix. Such efforts are necessary in order to uphold the
standard model and to construct the more fundamental theory beyond it. In
this note we propose a new phenomenological formula for the KM matrix which
has an interesting geometric picture and study a possible mechanism for its
origin in the standard model.\par

  Let us consider a three dimensional representation space for the quark
masses and introduce in it the two unit vectors as
 \begin{equation}
   \be^{(u)}=\frac{1}{\sqrt{m_u^2+m_c^2+m_t^2} }
                           \left(
                           \begin{array}{c}
                         m_u\\      
                         m_c\\
                         m_t
                           \end{array}
                           \right),\quad 
   \be^{(d)}=\frac{1}{\sqrt{m_d^2+m_s^2+m_b^2 }}
                           \left(
                           \begin{array}{c}
                         m_d\\
                         m_s\\
                         m_b
                           \end{array}
                           \right)
 \end{equation}
for the up and down quark sectors. Our basic postulate is to interpret one
of unitary matrices, $V$, satisfying
 \begin{equation}
   \be^{(u)}=V \be^{(d)}
   \label{def}
 \end{equation}
as the KM matrix. \par

  An orthogonal matrix which transforms $\be^{(d)}$ to $\be^{(u)}$ is 
readily constructed to be
 \begin{eqnarray}
     \nonumber 
   T &=& {\rm e}^{-\fbn\cdot\fbJ \theta}\\
     \nonumber 
     &=& I \cos\theta - \bn\cdot\bJ\sin\theta + (1-\cos\theta)\,\bn\,{}^t\bn\\
     &=& I - \bn\cdot\bJ\sin\theta + (1-\cos\theta)(\bn\cdot\bJ)^2
 \label{ortho}
 \end{eqnarray}
where $\bn$ is the unit vector perpendicular to $\be^{(\alpha)}$ 
($\alpha = u, d$) as
 \begin{equation}
   \bn = {1 \over \sin\theta}\,{\be^{(u)}\times\be^{(d)}},
   \quad \be^{(u)}\cdot\be^{(d)} = \cos\theta
 \end{equation}
and
 \begin{equation}
   J_{1}=\left(
     \begin{array}{ccc}
       0 & 0 & 0 \\
       0 & 0 &-1 \\
       0 & 1 & 0
     \end{array}
   \right), \quad
   J_{2}=\left(
     \begin{array}{ccc}
       0 & 0 & 1 \\
       0 & 0 & 0 \\
      -1 & 0 & 0
     \end{array}
   \right), \quad
   J_{3}=\left(
     \begin{array}{ccc}
       0 &-1 & 0 \\
       1 & 0 & 0 \\
       0 & 0 & 0
     \end{array}
   \right)
 \end{equation}
are generators of the SO(3) group satisfying
 \begin{equation}
   [J_{i},J_{j}] = \varepsilon_{ijk}J_{k}.
 \end{equation}
To calculate quantities including $\bJ$, it is convenient to use the following
formulas for arbitrary vectors $\ba$ and $\bb$ as
 \begin{equation}
   (\ba\cdot\bJ)\,\bb = \ba\times\bb,
 \end{equation}
 \begin{equation}
   [\ba\cdot\bJ,\,\bb\cdot\bJ] = (\ba\times\bb)\cdot\bJ
 \end{equation}
and
 \begin{equation}
   (\ba\cdot\bJ)(\bb\cdot\bJ) = - (\ba\cdot\bb)\,I + \bb\,{}^t\ba.
 \end{equation}
The rotation around each axis $\be^{(\alpha)}$ ($\alpha = u,\,d$) leaves the
axis unchanged. Namely, the vector $\be^{(\alpha)}$ is invariant under the
orthogonal transformation
 \begin{eqnarray}
   \nonumber
   R_\alpha(\phi_\alpha)
   & = &{\rm e}^{-\fbea\cdot\fbJ\phi_\alpha}\\
   \nonumber
   & = & I\cos \phi_\alpha -\be^{(\alpha)}\cdot\bJ\sin\phi_\alpha
     +   (1-\cos\phi_\alpha)\be^{(\alpha)}\,{}^t\be^{(\alpha)}\\
   & = & I - \be^{(\alpha)}\cdot\bJ\sin\phi_\alpha
     +   (1-\cos\phi_\alpha)(\be^{(\alpha)}\cdot\bJ)^2
 \end{eqnarray}
for an arbitrary angle $\phi_\alpha$. Therefore, we have a class of
orthogonal matrices consisting of $R_u(\phi_u)$, $T$ and $R_d(\phi_d)$ with
arbitrary angles $\phi_u$ and $\phi_d$ that transforms $\be^{(d)}$ to
$\be^{(u)}$.

  It is crucial to note that there exists another class of transformations
which leaves Eq.(\ref{def}) unaltered. Namely, Eq.(\ref{def}) is invariant
under one parameter U(1) operation generated by a matrix whose kernel consists
of the vectors $\be^{(u)}$ and $\be^{(d)}$.  Such a matrix is found to be
$\bn\,{}^t\bn = I + (\bn\cdot\bJ)^2$. In fact both vectors $\be^{(u)}$ and
$\be^{(d)}$ are invariant under the action of the unitary matrix
 \begin{eqnarray} 
              \nonumber
   U(\delta) & = & {\rm e}^{i\delta\fbn\,{}^t\fbn}\\
              \nonumber
             & = & I + \left({\rm e}^{i\delta} - 1 \right)\bn\,{}^t\bn\\
             & = & {\rm e}^{i\delta}I
                   + \left({\rm e}^{i\delta} - 1 \right)(\bn\cdot\bJ)^2
 \end{eqnarray}
for an arbitrary phase $\delta$, since
 \begin{equation}
   (\bn\,{}^t\bn)\be^{(\alpha)}=0\quad (\alpha = u,\,d).
 \end{equation}
Combining the redundant $R_\alpha(\phi_\alpha)$ and $U(\delta_\alpha)$
operations with $T$, we get a class of unitary transformations $V$
satisfying Eq.(\ref{def}).\par

  It is natural to interpret that the redundancy related to the
$R_\alpha(\phi_\alpha)$ and $U(\delta_\alpha)$ transformations is inherent
in the two vectors $\be^{(u)}$ and $\be^{(d)}$ in the representation space
for the quark masses. The vector $\be^{(\alpha)}$ is indistinguishable from
the class of vectors
$U^\dagger(\delta_\alpha){}^tR_\alpha(\phi_\alpha)\be^{(\alpha)}$ with
arbitrary angle $\phi_\alpha$ and phase $\delta_\alpha$.
Accordingly, Eq.(\ref{def}) is essentially identical with
 \begin{equation}
  \be^{(u)}
  = R_u(\phi_u)U(\delta_u)\,T\,U^\dagger(\delta_d)\,{}^tR_d(\phi_d)\,\be^{(d)}
  \label{equivdef}
 \end{equation}
for arbitrary $\delta_\alpha$ and $\phi_\alpha$. Therefore, the orthogonal
matrix $T$ in Eq.(\ref{ortho}) and the unitary matrix
 \begin{equation}
   V(\phi_u,\,\phi_d,\,\delta) = R_u(\phi_u)\,T\,U(\delta)\,{}^t R_d(\phi_d),
   \quad   \delta \equiv \delta_u-\delta_d
 \label{VKM}
 \end{equation}
are equivalent with respect to the action on the vector $\be^{(d)}$ in the
quark mass space. However, $T$ and $V$ acquire different physical meanings
as the transformation matrices when they are postulated to act on a vector
consisting of the chiral quark fields in the three-dimensional generation
space. It is the unitary matrix $V(\phi_u,\,\phi_d,\,\delta)$ endowed with
appropriate values for the angles $\phi_\alpha$ and phase $\delta$ that is
interpreted as the KM matrix in this article. \par

 Owing to the identity
 \begin{equation}
   {\rm e}^{-\fbn\cdot\fbJ\theta}\be^{(d)}\cdot\bJ
   = \be^{(u)}\cdot\bJ{\rm e}^{-\fbn\cdot\fbJ\theta},
 \label{ndun}
 \end{equation}
the unitary matrix $V$ has the other expressions as
 \begin{equation}
   V(\phi_u,\,\phi_d,\,\delta) = T\,R_d(\phi_u)\,U(\delta)\,{}^tR_d(\phi_d)
                               = R_u(\phi_u)\,U(\delta)\,{}^tR_u(\phi_d)\,T.
   \label{trdr}
 \end{equation}
This matrix function satisfies the periodic property
 \begin{equation}
  V(\phi_u+m\pi,\,\phi_d+n\pi,\,\delta) = V(\phi_u,\,\phi_d,\,\delta)
 \label{period} 
 \end{equation}
for integers $m$ and $n$ provided that $m + n =$ even. To prove
Eq.(\ref{period}), it is sufficient to show the relation
 \begin{eqnarray}
   R_d(\phi_u)\,U(\delta)\,{}^t R_d(\phi_d)
    &=&  (I\cos \phi_u -\be^{(d)}\cdot \bJ\sin\phi_u
         - \cos\phi_u\be^{(d)}\,{}^t\be^{(d)})\,U(\delta)
   \nonumber \\
   \noalign{\vskip 0.2cm}
    & &\times (I\cos \phi_d -\be^{(d)}\cdot \bJ\sin\phi_d
         - \cos\phi_d\be^{(d)}\,{}^t \be^{(d)})
   \nonumber \\
   \noalign{\vskip 0.2cm}
    & &+ \be^{(d)}\,{}^t\be^{(d)}
 \end{eqnarray}
by using the identities
$\be^{(d)}\,{}^t \be^{(d)}\,U(\delta) = U(\delta)\,\be^{(d)}\,{}^t \be^{(d)}
                                      = \be^{(d)}\,{}^t \be^{(d)}$.\par

  At present the world averages of the absolute values of KM matrix elements
are estimated as follows [10]:
 \begin{equation}
   \left(
     \begin{array}{ccc}
       0.9747\sim0.9759 & 0.218\sim0.224   & 0.002\sim0.005 \\
       0.218\sim0.224   & 0.9738\sim0.9752 & 0.032\sim0.048 \\
       0.004\sim0.015   & 0.030\sim0.048   & 0.9988\sim0.9995
     \end{array}
   \right).
   \label{expdata}
 \end{equation}
Numerical estimation of our formula for the KM matrix requires the masses of
six quarks at the same energy scale. Using the 2-loop renormalization group,
Koide [11] obtained the quark masses at the energy
scale $\mu=1 {\rm GeV}$ as
 \begin{equation}
  \begin{array}{ll}
    m_u = 5.6\pm1.1\ {\rm MeV},\quad  & m_d = 9.9\pm1.1\ {\rm MeV},\\
    m_s = 199\pm33\ {\rm MeV},\quad   & m_c = 1492^{+23}_{-40}\ {\rm MeV},\\
    m_b = 7005^{+29}_{-53}\ {\rm MeV},\quad & m_t = 424^{+54}_{-66}\ {\rm GeV}.
  \end{array}
 \end{equation}

For the sake of definiteness, we use the following values
 \begin{equation}
  \begin{array}{ll}
   m_u = 5.6\ {\rm MeV},\quad  & m_d = 9.9\ {\rm MeV},\\
   m_s = 202\ {\rm MeV},\quad  & m_c = 1492\ {\rm MeV},\\
   m_b = 7005\ {\rm MeV},\quad & m_t = 400\ {\rm GeV} 
  \end{array}
 \end{equation}
for the quark masses in the present analysis. In the case of $\delta=0$,
the unitary matrix $V$ in Eq.(\ref{trdr}) is reduced into the one parameter
form
 \begin{equation}
   V = T\,R_d(\phi) = R_u(\phi)\,T,\quad \phi \equiv \phi_u-\phi_d.
 \end{equation}
The least-square fitting determines the KM matrix elements to be
 \begin{equation}
   V=\left(
     \begin{array}{ccc}
       -0.97526  & 0.22100  & -0.0049805 \\
       -0.22104  & -0.97473 & 0.032152 \\
       0.0022508 & 0.032457 & 0.99947
     \end{array}
   \right),
   \label{delta0}
 \end{equation}
by choosing
 \begin{equation}
   \phi = -3.3645.
\end{equation}
The absolute values of Eq.(\ref{delta0}) are in good agreement with
experimental data in Eq.(\ref{expdata}). For $\delta \ne 0$, we find
several sets of values for the parameters $(\phi_u,\,\phi_d,\,\delta)$
which constrain the KM matrix elements within the experimental
uncertainties.  The best choice is the following set as
 \begin{equation}
  (\phi_u,\,\phi_d,\,\delta) = (0,\,3.32918,\,{\pi \over 3})
   \label{bestfit}
 \end{equation}
which leads to
 \begin{equation}
   V=\left(
     \begin{array}{lll}
      0.97583\,{\rm e}^{-2.1063\,i} & 0.21849\,{\rm e}^{1.2629\,i}
                              & 0.0049633\,{\rm e}^{-1.8133\,i}\\
      0.21853\,{\rm e}^{2.9259\,i}  & 0.97530\,{\rm e}^{-3.1297\,i}
                              & 0.032157\,{\rm e}^{0.0083506\,i}\\
      0.0022219\,{\rm e}^{-0.073299\,i} & 0.032461\,{\rm e}^{0.0012436\,i}
                              & 0.99947\,{\rm e}^{-0.000001\,i}
     \end{array}
   \right).
 \end{equation}
Exceptionally the $V_{31}$ component tends to have a somewhat small value
and the value of $|V_{ub}|/|V_{cb}|$ is somewhat large compared with
experimental results. In the case of $\delta = {\pi\over3}$, the
rephasing-invariant Jarlskog parameter [12--16] 
$J={\rm Im}(V_{23}V_{12}V_{22}^\dagger V_{13}^\dagger)$ moves in the range
 \begin{equation}
   - 5.9\times10^{-5} \le J(\phi_u,\phi_d,{\pi \over 3})
                      \le 5.9\times10^{-5}.
 \end{equation}
For the parameter set in Eq.(\ref{bestfit}), $J$ takes the value
 \begin{equation}
   J = - 2.3437\times 10^{-6},
\end{equation}
the magnitude of which is somewhat small in comparison with the present
data [17,18]. The parameter set giving a larger value for $|J|$
tends to deviate the KM matrix elements calculated by Eq.(\ref{VKM}) from
the experimental data. In these results, there are some diviations from
experiment. We consider they are probably due to the way of choosing the
quark mass values. \par

  In the standard model the Higgs mechanism generates the fermion masses in
the forms of mass matrices and the KM matrix is given by the product of
unitary matrices diagonalizing the up and down quark mass matrices.  In
this note the KM matrix is derived as a unitary matrix relating the mass
vectors in the generation space. To give a theoretical foundation to our
formula for the KM matrix in Eq.(\ref{VKM}), we must reformulate the mass
vector description in the ordinary mass matrix scheme of the standard
model. In fact it is possible to show that non-hermitian quark mass
matrices being diagonalizable by a unitary matrix and a unit matrix lead
to the mass vector description under a simple additional condition. \par

  Let us express the chiral component fields of the up and down quarks
generically by $u_h^i$ and $d_h^i$ ($h = L,\,R$ ; $i = 1,\,2,\,3$).
The hermitian Lagrangian density for fermion masses is given by 
 \begin{equation}
   {\cal L}_{\rm mass}=
           \sum_{i,j}^3\,(\,{\bar u}_L^i M^u_{ij} u_R^j
                           + {\bar d}_L^i M^d_{ij} d_R^j
                           + {\bar u}_R^j M^{u*}_{ij} u_L^i
                           + {\bar d}_R^j M^{d*}_{ij} d_L^i \,)
 \end{equation}
with the mass matrices
 \begin{equation}
   M_\alpha = {v \over \sqrt{2}}\left( Y^\alpha_{ij} \right),
   \quad(\alpha=u,\,d)
   \label{mmatrix}
 \end{equation}
where $Y_{ij}^\alpha\,(\alpha=u,d)$ are the Yukawa coupling constants
and $v$ is a vacuum expectation of the Higgs field. \par

  Here we postulate first that the mass matrices $M_\alpha$ are non-hermitian
and diagonalized as follows :
 \begin{equation}
   M_\alpha\ \rightarrow\ U_L^\alpha M_\alpha U_R^{\alpha\dagger} =
   U_L^\alpha M_\alpha I = M^{diag}_\alpha = ( m_i^\alpha \delta_{ij} ),
   \quad(\alpha=u,\,d)
   \label{diagonal}
 \end{equation}
by perfectly-asymmetric bi-unitary transformations
 \begin{equation}
  \left\{
  \begin{array}{l}
    u_L^i \rightarrow (U_L^u)^i_j u_L^j\\
   \noalign{\vskip 0.2cm}
    u_R^i \rightarrow (U_R^u)^i_j u_R^j = u_R^i,
  \end{array}
  \right.
 \qquad
  \left\{
  \begin{array}{l}
    d_L^i \rightarrow (U_L^d)^i_j d_L^j\\
   \noalign{\vskip 0.2cm}
    d_R^i \rightarrow (U_R^u)^i_j d_R^j = d_R^i.
  \end{array}
  \right.
 \end{equation}
The generation space vectors $u_L = (u_L^i)$ and $d_L =(d_L^i)$ consisting
of the left-handed chiral fields of electroweak doublets receive unitary
transformations $U_L^u$ and $U_L^d$. In contrast the vectors $u_R = (u_R^i)$
and $d_R = (d_R^i)$ of the right-handed chiral fields of electroweak singlets
remain as eigenstates in both the interaction and mass eigenmodes.  As the
second postulate we require that the action of the mass matrices $M_u$ and
$M_d$ transform the vector
 \begin{equation}
     \be = {1 \over \sqrt{3}}
           \left( \begin{array}{c}
                    1\\
                    1\\
                    1
                  \end{array}
           \right)
     \label{evector}
 \end{equation}
in the three dimensional generation space into parallel vectors, {\it viz.},
 \begin{equation}
     M_u \be\ \parallel\ M_d \be
   \label{parallel}
 \end{equation}
which results in
 \begin{equation}
   U^{u\dagger}_L
    \left( \begin{array}{c}
             m_u \\
             m_c \\
             m_t
           \end{array} \right)
     \ \parallel\ 
   U^{d\dagger}_L
    \left( \begin{array}{c}
             m_d \\
             m_s \\
             m_b
           \end{array} \right) 
   \label{ }
 \end{equation}
owing to Eq.(\ref{diagonal}). Then we get
 \begin{equation}
   \be^{(u)} = U_L^u U_L^{d\dagger} \be^{(d)}
 \end{equation}
which should be identified with the relation in Eq.(\ref{equivdef}).\par

  In this way we have found a mechanism which relates the mass vector
description to the ordinary mass matrix formalism in the standard model.
For such a mechanism to work, the quark mass matrices must have the form
\[
    M_\alpha = U_L^{\alpha\dagger}M_\alpha^{diag},\quad (\alpha = u,\, d)
\]
and the generation vector space must have a direction of anisotropy specified
by the $\be$ vector in Eq.(\ref{evector})\footnote{If the present formula 
is applied to the lepton sector, the neutrinos must necessarily be massive.}.
In this connection it is worthwhile to mention that Foot [19] 
introduced a generation space where the vectors consist of the square roots
of quark masses and the $\be$ vector plays a special role. In such a vector
space he found an interesting geometrical interpretation for Koide's lepton
mass formula [4,20]. Esposito and Santorelli [21] extended
the Foot method to the quark masses and the Dirac masses of neutrinos. \par

  One of us (I.\,S.\,S.) would like to express his sincere thanks to
Professor T. Maskawa for discussion and comment.

\end{document}